\documentclass[sigchi, screen, authorversion]{acmart}

\setcopyright{none}
\renewcommand\footnotetextcopyrightpermission[1]{} 
\usepackage{graphicx}
\usepackage{subcaption}
\usepackage{float}

\graphicspath{{figures/}}
\usepackage{xspace}
\newcommand{\ie}{{i.e.}\xspace}
\newcommand{\eg}{{e.g.,}\xspace}

\newcommand{\ea}{{et~al.}\xspace}

\newcommand{\etc}{{etc.}\xspace}

\usepackage{tikz}

\begin{document}

\title{Visualizing a Million Time Series with \\the Density Line Chart}

\renewcommand{\shortauthors}{D. Moritz et al.}

\author{Dominik Moritz}
\affiliation{\institution{Paul G. Allen School of Computer Science \& Engineering, University of Washington}}
\email{domoritz@cs.washington.edu}
\authornote{This work was started at Microsoft Research.}

\author{Danyel Fisher}
\affiliation{\institution{Honeycomb}}
\email{danyel@gmail.com}
\authornotemark[1]

\begin{abstract}
Data analysts often need to work with multiple series of data---conventionally shown as line charts---at once. Few visual representations allow analysts to view many lines simultaneously without becoming overwhelming or cluttered. In this paper, we introduce the DenseLines technique to calculate a discrete density representation of time series. DenseLines normalizes time series by the arc length to compute accurate densities. The derived density visualization allows users both to see the aggregate trends of multiple series and to identify anomalous extrema.
\end{abstract}



 \begin{teaserfigure}
   \includegraphics[width=\textwidth]{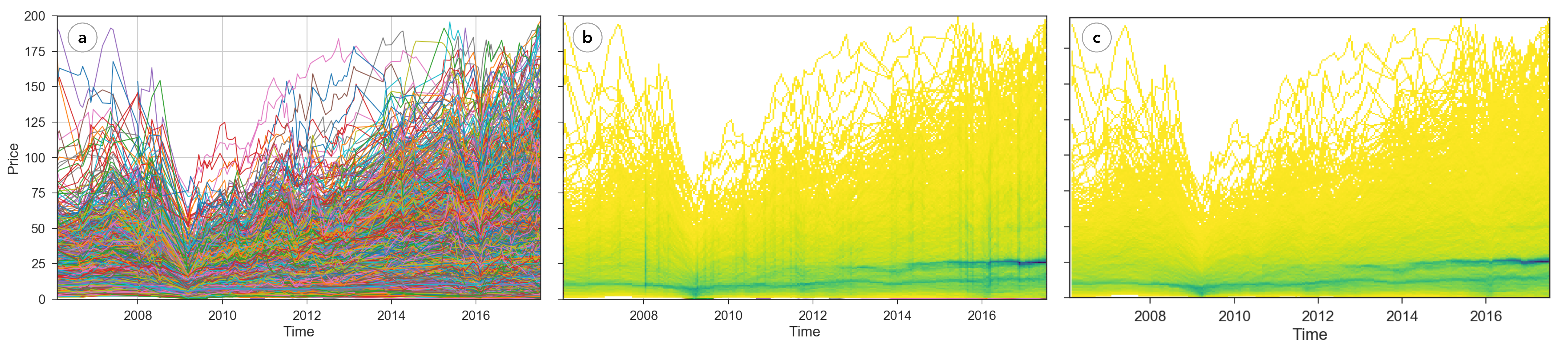}
   \vspace{-16px}
   \caption{Line chart (a), non-normalized heatmap (b), and DenseLines heatmap (c). The line chart suffers from overdraw and clutter. The heatmap without normalization has visible artifacts (vertical lines). In DenseLines, the intensity of the color shows the density of lines. Note the visible high points across the top, the collective dip in stocks during the crash of 2008, and the two distinct bands of \$25 and \$15 stocks.}
   \label{fig:teaser}
 \end{teaserfigure}

\maketitle

\section{Introduction}

Time series are a common form of recorded data, in which values continuously change over time but must be measured and sampled at discrete intervals. Time series play a central role in many domains~\cite{Fulcher2013}: finance and economics (stock data, inflation rate), weather forecasting (temperature, precipitation, wind, pollution), science (radiation, power consumption), health (blood pressure, mood levels), and public policy (unemployment rate, income rate) to name a few. 
Often, an individual time series corresponds to a context such as the location of a sensor.
Therefore, analysts may have many series to consider---multiple stocks or the unemployment rates in different counties. These multiple contexts can result in datasets with as many as thousands of time series.


Multiple series are typically visualized as line charts with one line per series~\cite{playfair1801commercial}. However, even with as few as a hundred lines, overplotting makes it difficult for analysts to see patterns and trends. Existing techniques simply do not scale to these numbers of series. A na\"ive density-based technique suffers from a different issue: lines with extreme slopes are overrepresented in the visualization.

\begin{figure*}[t]
  \centering
  \includegraphics[width=\textwidth]{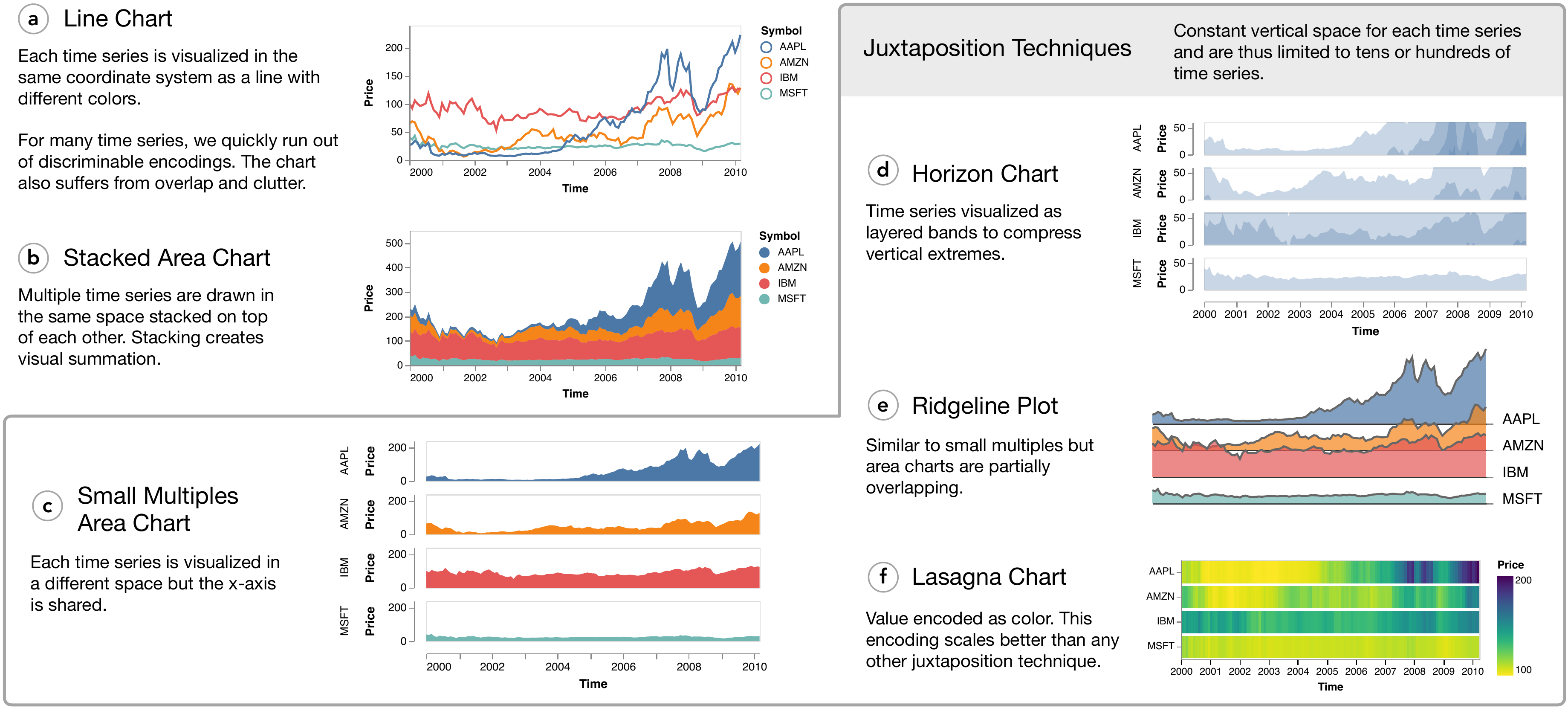}
  \vspace{-10px}
  \caption{Variants of time series visualizations for the same four stock prices over time. In all visualization types, with more time series the visual clutter increases or more vertical space is required.}\label{fig:overview}
\end{figure*}

We present the DenseLines technique, which allows analysts to make sense of many time series. In this paper, we show that the technique is scalable in the number of series, at the cost of removing the ability to trace individual lines. DenseLines allows analysts to answer questions such as: ``What are the major trends in my time series data?'' and ``Are these time series behaving similarly to each other?'' The core of the technique is to compute a density as the number of time series that pass through a particular space in the time and value dimensions; and to  normalize the density contribution of each line by its arc length, such that each series has the same total weight. The density can then be visualized with a color scale, as seen in~\autoref{fig:teaser} (c). The technique is scalable, meaning that additional lines or higher resolution data do not affect the visual complexity of the chart; it is amenable to interaction techniques and different color scales.

We validate the technique through a series of examples, including stock data, hard drive statistics, a case study of data analysts at a large cloud services organization, and with a synthetic dataset of a million time series. Our implementations of DenseLines in Rust\footnote{\url{https://github.com/domoritz/line-density-rust}} and in JavaScript with WebGL\footnote{\url{https://domoritz.github.io/line-density}} are available as open source.

\section{Related Literature}

The standard encoding of time series---time mapped to a horizontal axis and value to the vertical axis, with line segments connecting the points---has been in use for centuries~\cite{playfair1801commercial}. Multiple series can be visualized as superimposed lines, each with a different color or other distinctive encodings (\autoref{fig:teaser} (a), \autoref{fig:overview} (a)).

\subsection{Visualizing Many Series}

Javed~\ea~\cite{Javed2010} survey visualization techniques for line charts with multiple time series. They empirically compare the design's effectiveness for varying tasks and numbers of series. One important finding is that clutter~\cite{clutter} can be overwhelming to users: presenting users with more lines tends to decrease correctness in perceptual tasks while also increasing task completion time. Even for fairly small numbers of series---Javed~\ea limit themselves to eight while previous studies were often restricted to two~\cite{Heer2009, Simkin1987}---chart elements rapidly lose discriminability and become cluttered. 

Juxtaposition~\cite{6183556}---placing charts next to each other---reduces clutter but requires more space (\autoref{fig:overview} c - f)). LiveRAC~\cite{McLachlan2008} uses a matrix of reorderable small multiples (\autoref{fig:overview} (c))~\cite{tufte1985visual} to provide high information density for exploring larger numbers of time series. Horizon charts (\autoref{fig:overview} (d))~\cite{Saito2005, Heer2009} reduce the space in charts by dividing the line into layered bands. Ridgeline plots (\autoref{fig:overview} (e)) instead allow overlap between the time series\footnote{This representation is inspired by the classic 1979 Joy Division ``Unknown Pleasures'' album cover. It shows a figure from the PhD thesis of the astronomer Harold Craft, who recorded radio intensities of the first known pulsar~\cite{craft1970radio}.}.

A time series can also save space by encoding value as color, and so use a small, but constant, amount of vertical space (\autoref{fig:overview} (f)). Swihart~\ea coined the term ``Lasagna Plot''~\cite{Swihart2010} for this representation to contrast it with the line chart with too many lines. Rather than having tangled ``noodles'' (lines), each series is shown as a layer through time. The Line Graph Explorer~\cite{Kincaid2006} uses this technique to enable users to explore dozens of time series. Juxtaposition maintains an ability to look at each of the series, and is so limited in the degree to which it can scale. It is thus useful for a small number of series; on the order of tens or at most hundreds.

\autoref{fig:overview} compares time series visualizations but we find that ultimately none scale to visualizing large numbers of time series at the same time. A broad selection of visual designs found in~\cite{aigner:vistime:2011} build on these patterns and share the limitations.

Each visualization technique emphasizes different properties of the data and are thus preferred in particular domains. For example, neuroscientists often use ridgeline plot because they care about seeing where high peaks occur~\cite{}. In juxtaposed visualizations the order matters and time series that are close are easier to compare than those that are far apart. DenseLines plot all data in the same space and emphasizes density and outliers. 

\subsection{Searching for Specific Patterns or Insights}

Rather than attempting to visualize all the series, another approach is to search the dataset for lines that behave in particular ways. Wattenberg's QuerySketch~\cite{Wattenberg:2001:SGQ:634067.634292} and Hochheiser and Shneiderman's TimeBoxes~\cite{Hochheiser2004} allow users to select a subset of lines based on their shape characteristics. These techniques scale to very large sets of time series but provide a limited view of the data. Konyha~\ea discuss interaction techniques for filtering time series data~\cite{Konyha2012}.

\subsection{Visualizing Density}

The design of DenseLines draws its inspiration from density visualizations, which are commonly used to declutter scatterplots~\cite{Carr1987}. Density alone is sufficient to see trends, clusters, and other patterns, and to recognize outlier regions~\cite{Wickham2013}. Past work has plotted density marks by reducing the opacity of the marks~\cite{Hinrichs2015}, by smoothing~\cite{Wickham2013}, or by binning data across both the X and Y values, and then encoding the number of records in each bin using color. Compared to bagplots and boxplots for time series data~\cite{functionalbagplots}, density based visualizations do not merge different groups in multi-modal data (\eg bundles of similar time series). A density representation can also be applied to other chart types such as network graphs~\cite{zinsmaier2012interactive}, and trajectories~\cite{Scheepens}. Continuous Parallel Coordinates~\cite{Heinrich:2009:CPC:1638611.1639151} and Parallel Coordinates Density Plots~\cite{PCDensity} visualize parallel coordinate plots for high dimensional data with many records. Parallel Edge Splatting~\cite{Burch:2011:PES:2068462.2068635} visualizes networks that evolve over time, and uses the increased density of line crossings to show how subsequent generations of the network differ.  

With hundreds or thousands of time series it becomes less important to trace individual lines. Analysts often want to know the amount of data in regions of a particular time and value. Visualization designers often use transparency blending methods. However, similar to transparency blending in scatterplots, there are two main drawbacks. If the opacity is set too low, individual outlier lines may become invisible. If the opacity is set too high, dense regions with different densities become indistinguishable. Heatmaps are a widely used, scalable alternative to scatterplots that address this issue by explicitly mapping the density in a discretized space to color. DenseLines follows this basic pattern and provides a scalable alternative to line charts by counting the amount of data in regions of a particular time and value.

\begin{figure}[b]
  \centering
  \includegraphics[width=\columnwidth]{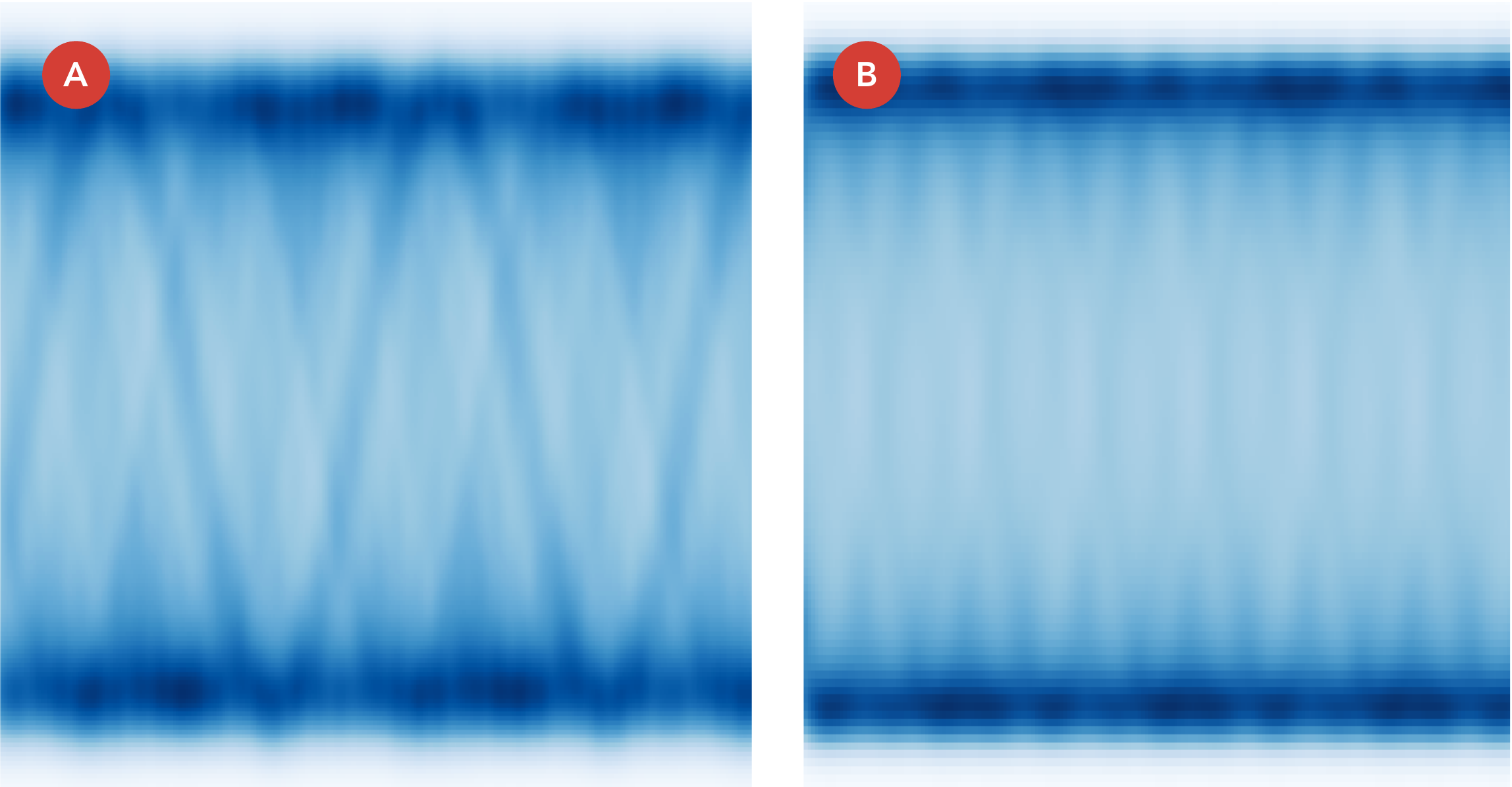}
  \vspace{-10px}
  \caption{Comparison of running Curve Density Estimates for 1000 time series (left) and DenseLines with a post-processing step to smooth the densities with a Gaussian kernel (right). DenseLines is multiple orders of magnitude faster.}~\label{fig:binsummarize}
  \vspace{-10px}
\end{figure}

Lampe and Hauser~\cite{Lampe2011} proposed Curve Density Estimates, which uses kernel density estimation to render smooth curves. DenseLines is a special case of Curve Density estimates where data is aggregated into bins; the output is discrete rather than smooth. DenseLines is to Curve Density Estimates what discrete histograms (binned plots) are to smoothed density estimates. They share similar disadvantages and advantages. One the one hand, excessive variability in aggregates of a binned plot can distract from the underlying pattern. On the other hand, smoothing can ``smear'' values into areas without data---if the count in a cell in a binned plot is more than zero there must be data in the cell. While smooth summaries can be statistically more robust, binned summaries are easier to compute. DenseLines can be computed faster than Curve Density Estimates and are also easier to implement, which could help with the adoption of the technique. For large datasets, we can approximate smooth density estimates without sacrificing performance. For this, we can follow the Bin-Summarize-Smooth approach by Hadley Wickham~\cite{Wickham2013} and bin and summarize with DenseLines first and then smooth the output. By smoothing the summarized output---whose size only depends on the resolution but not the original data---we can compute output similar to Curve Density Estimates for large data in a fraction of the time (\autoref{fig:binsummarize}).

\subsection{Arc Length in Data Visualization}

In DenseLines, we normalize the contribution of a line to the density by the arc length. This normalization precisely corrects the additional ink of steep slopes. After normalization, each time series contributes equally to the heatmap. In a regular line chart, the average value has to be computed by sampling values at regular intervals along the x-axis. In a normalized line chart, the average is the weighted average of regular samples along the line itself. A ``normalized line chart'' might thus aid in aggregate tasks over time series data similar to colofields~\cite{comparingaverages}. Scheidegger~\ea~\cite{Scheideggernormalize} normalized properties of isosurfaces to derive statistics of the underlying data using a similar method. Normalization for time series data makes similar assumptions and has similar goals as the mass conservation method Heinrich and Weiskopf use in Continuous Parallel Coordinates~\cite{Heinrich:2009:CPC:1638611.1639151}. Talbot~\ea~\cite{Talbot2011ArcLA} use arc length to select a good aspect ratio in charts. However, we are the first to use it to normalize line charts. The normalization yields similar results as the column-normalized grids in Lampe~\ea's Curve Density Estimates~\cite{Lampe2011} but does not rely on a kernel to compute densities. 

\section{The Design of DenseLines}

A chart representing multiple time series may support a number of different tasks. Our goal is to let analysts recognize dense regions along both the time and the value dimension while preserving extrema. These tasks represent user tasks that are common for telemetry data from monitoring clusters of servers: analysts have an interest in knowing about collective user behavior and server performance. In addition to supporting these tasks, the representation should scale: additional time series should not impede interpretation.

The DenseLines technique focuses on the visualization of multiple time series with identical temporal and similar value domain. Similar to multi-series line chart, DenseLines uses unified chart axes.
However, rather than showing individual series, our goal is to support the analysis of dense areas in the chart (local areas where many time series pass), as well as extreme values (outliers).
A DenseLines chart defines local density as the density of lines. We compute density by binning the chart into regions; the density of a bin measures the number of lines that pass through that bin. This definition is more subtle than for a scatterplot heatmap: the data that underlies line charts is not usually recorded at every point. Rather, a line chart connects a set of time/value pairs; the technique must count how many different series lines pass through each bin.

\subsection{Normalization of Density by Arc Length}

\begin{figure}[b]
  \centering
  \includegraphics[width=\columnwidth]{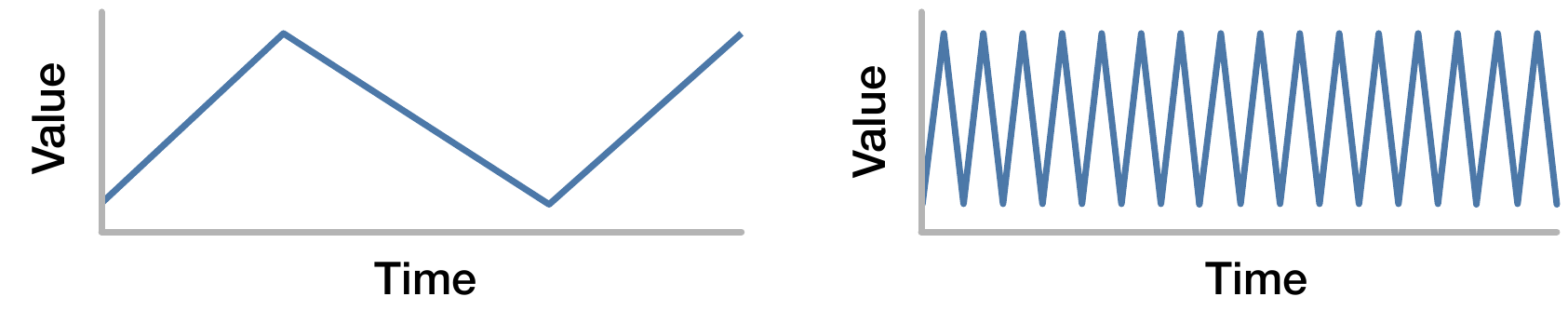}
  \vspace{-10px}
  \caption{Two line charts that span the same time and have the same average value. The right series has more variability, which leads to more pixels drawn for the same amount of data.}~\label{fig:freq}
  \vspace{-10px}
\end{figure}

\begin{figure}[b]
  \centering
  \vspace{-10px}
  \includegraphics[width=\columnwidth]{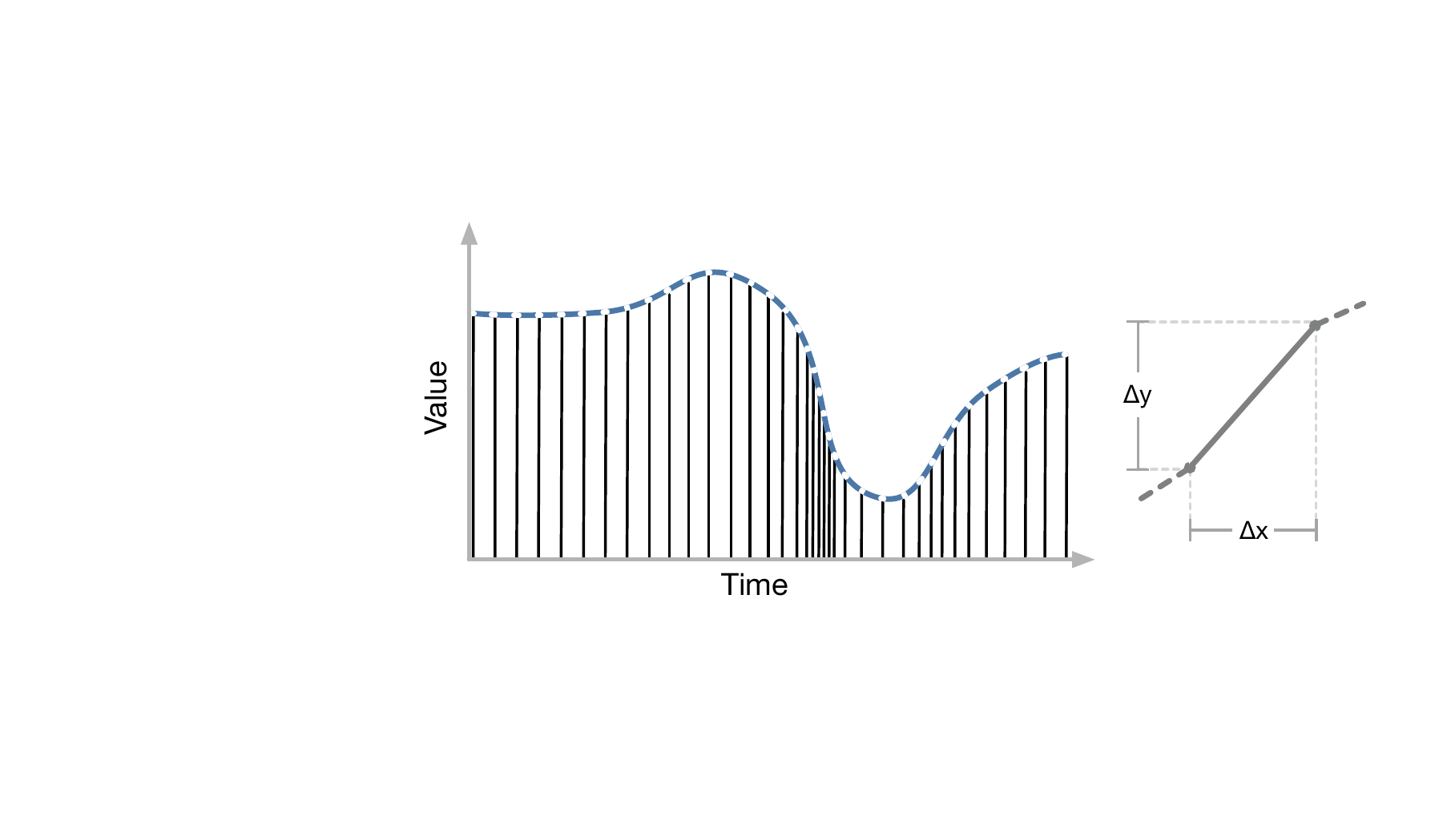}
  \vspace{-20px}
  \caption{With uniform sampling along a line, steep segments are denser when projected onto the time axis. To use the same weight for each segment of the same length in time, we need to normalize by the the arc length.}~\label{fig:sample}
\end{figure}

Line charts present a distinct challenge: lines with steep slopes are rendered with more pixels. Since a time series is a continuous value that is recorded at discrete intervals, both time series in \autoref{fig:freq} can be defined by the same number of data points. Both series have the same average value (and so the same area under the curve). However, the series on the right is plotted with more pixels. Consequently, density based techniques for time series give more weight to lines with steep slopes. \autoref{fig:sample} (left) shows that when the slope is steep, more points are needed for the same time span. We need to reduce the weight (\ie number of pixels or amount of ink) of steep line segments such that each line contributes equally to the density in the heatmap. Concretely, for any time span, the contribution of each series to the heatmap has to be the same.

We address this issue by normalizing the density of a line at a particular time by its arc length at that point in time. 
To understand why normalization by the arc length satisfies the requirements from above, we can look at a single line segment (\autoref{fig:sample}, right).
Within the same time interval, each series has the same extent in the time dimension but different extent in the value dimension.
To correct the contribution of each segment, we have to divide its weight by the length of the segment. Then every segment has the same weight regardless of its slope. The length of a segment with horizontal extent of $\Delta x$ and a vertical extent of $\Delta y$ can be derived from the Pythagorean theorem as $\sqrt{{\Delta x}^2 + {\Delta y}^2}$. In the limit of $\Delta x \to 0$ the length of an arc defined by $f$ and its first derivative (slope) $f'$ is $\sqrt{1+[f'(x)]^{2}}$. Notice that the difference between the arc length and the slope decreases with increasing slope. However, when the slope is $0$ (horizontal line), the normalization by arc length is $1$.



\subsubsection{Practical Approximation}

In DenseLines, we use a practical simplification for normalizing lines by arc length---as used in Curve Density Estimates by Lampe\~ea~\cite{Lampe2011}. In practice, we can assume one line segment per column (similar to the M4 time series aggregation~\cite{Jugel2014M4AV}) and normalize by the number of pixels drawn in each column. A horizontal line is not affected (normalization by~1). Also consistent with our requirements, every series gets the same overall weight. Mathematically, this simplification is asymptotically equivalent to normalization by arc length (in the limit of increasingly small bins).

\begin{figure*}[!htb]
  \centering
  \begin{subfigure}[t]{.32\textwidth}
    \includegraphics[height=80pt]{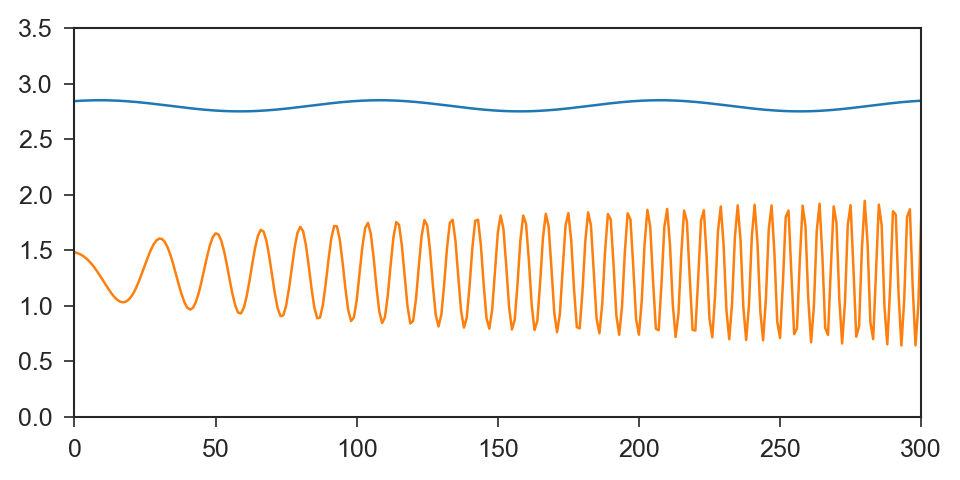}
    \caption{Model of two time series with a constant frequency (blue) and with increasing frequency and amplitude (orange).}~\label{fig:model_normalized}
  \end{subfigure} \hfill
  \begin{subfigure}[t]{.32\textwidth}
    \includegraphics[height=80pt]{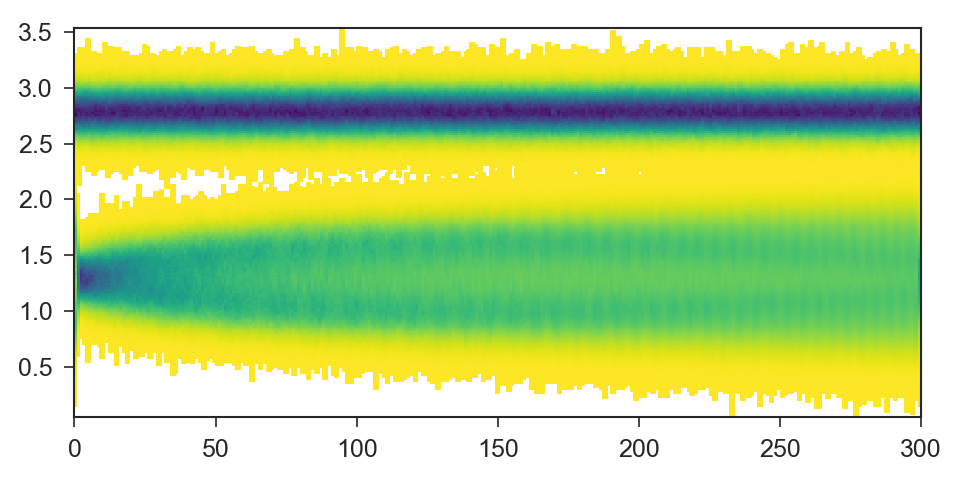}
    \caption{DenseLines of 10,000 time series sampled from the two time series in (a). The counts in each bin are normalized.}~\label{fig:normalized}
  \end{subfigure} \hfill
  \begin{subfigure}[t]{.32\textwidth}
    \includegraphics[height=80pt]{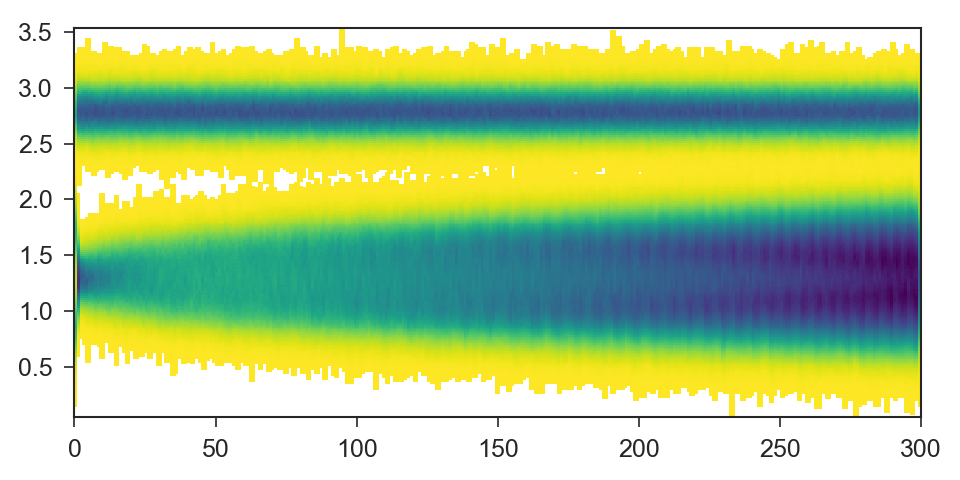}
    \caption{Visualization of counts without normalization. The density of the second group appears to increase to the right. Moreover, more time series appear to be generated from the orange line.}~\label{fig:not_normalized}
  \end{subfigure}
  \vspace{-10px}
  \caption{We use a model of with two time series (a) to generate 5,000 time series each. The 10,000 time series are visualized using DenseLines (b) and as a comparison without normalization (c).
  }\label{fig:normalize}
\end{figure*}


In a rasterized line chart, each column of a single line chart that is normalized by the arc length sums up to one. Lampe~\ea discuss in their Curve Density Estimates paper~\cite{Lampe2011} that this enables us to interpret each column as a 1D probability density estimate. A 1 indicates that all lines were 100\% of their time in the corresponding row. 0.5 shows that the lines combined spent 50\% in the row. For a DenseLines chart with many time series, the count in each cell is the number of lines that go through it but counting lines that are half of the time are in another cell (but the same column) only half \etc. With some explanation for new users, DenseLines charts can have meaningful color scales and legends.

\subsubsection{Problems of Density without Normalization}

A lack of normalization in DenseLines leads to visible artifacts (\autoref{fig:teaser} (b)) and can produce misleading results. To demonstrate this, we generated $10,000$ time series from a model of two time series \autoref{fig:model_normalized}. The first series is a sine wave with a constant frequency. The second series has a higher frequency. The frequency and the amplitude increase with time in the second series. \autoref{fig:normalized} shows density with normalization (DenseLines). It accurately shows constant density even as the frequency increases. The increasing amplitude is visible. Without the normalization (\autoref{fig:not_normalized}), the density of the second time series appears higher, although there are $5,000$ lines in each group. Moreover, the density appears to increase with time, which is also not true.

\subsection{DenseLines Algorithm}

\begin{figure*}[!htb]
  \centering
  \includegraphics[width=\textwidth]{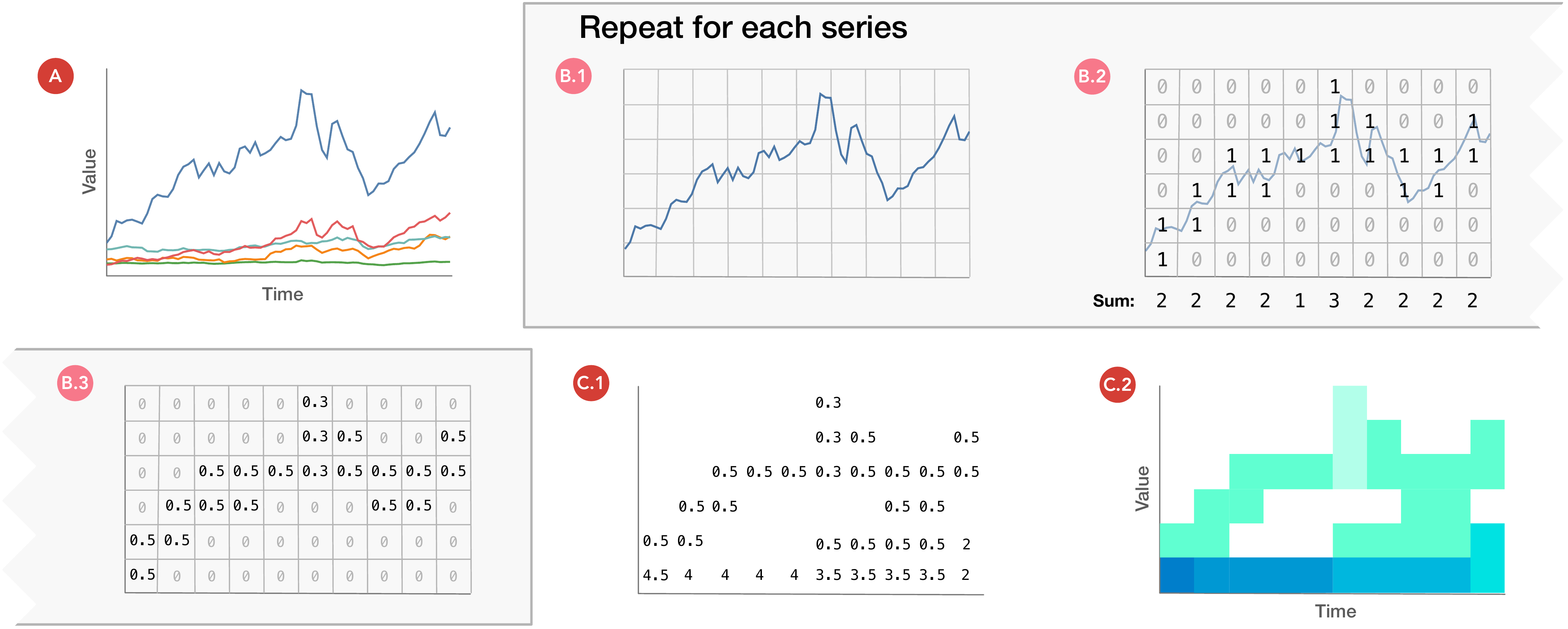}
  \vspace{-10px}
  \caption{The DenseLines algorithm for computing density for multiple time series has two steps. First, take a dataset of time series (A) and render each series in a discrete matrix (B.1). Set bins to 1 if the line passes through it (B.2). The matrix is then normalized by the sum in each column (B.3). In the second step, combine the normalized values into a single density map (C.1).}~\label{fig:algo}
  \vspace{-10px}
\end{figure*}

We compute the normalized density with the algorithm illustrated in \autoref{fig:algo}. The input is a dataset with many time series (A). We start by defining a two-dimensional matrix where each entry represents a bin in the time / value space. The time and value dimensions are discretized into equally sized bins. Using Bresenham's line algorithm~\cite{Bresenham1977}, we render the time series over the bins (B.1). Each bin that the line passes through gets a value of 1 and 0 otherwise (B.2). Alternatively, the value in each bin can correspond to the darkness of an anti-aliased line~\cite{antialias}.  We then normalize each value by the count of items in their column (B.3). These steps are repeated for every series. In a final step, the matrices of all time series are added together (C.1). The values in the matrix now represent the density of a particular bin in the time and value dimensions. The density can then be encoded using a color map (C.2).

Each line can be processed independently and the render, count, and normalize step can run in parallel. It can be implemented on MapReduce~\cite{Dean2004MapReduceSD} and on a GPU. We implemented DenseLines in a 
JavaScript Prototype with GPU computation\footnote{\url{https://github.com/domoritz/line-density}}.
Our implementation uses WebGL to processes $1\,\text{M}$ time series at a resolution of $400\text{px} \times 300\text{px}$ in \textasciitilde$40$ seconds on a 2014 MacBook Pro with Iris Graphics.
At a lower resolution of $32\text{px} \times 16\text{px}$, the algorithm runs for \textasciitilde$6$ seconds. Because the algorithm can be implemented efficiently on parallel processors and GPUs, densities can be recomputed at interactive speeds when the user wants to explore a subset of the time series, zoom, or change binning parameters.

We can tweak the scale that encodes the density values to emphasize certain patterns (\autoref{fig:algo}, (C.2)). For instance, by adding a discontinuity between zero density and the lowest non-zero density, we can ensure that outliers are not hidden~\cite{Kandel2012}. We can also apply smoothing~\cite{Wickham2013} to remove noise or run other analysis algorithms on the computed density map. For the examples in this paper we use Viridis~\cite{viridis}, a perceptually uniform, multi-hue color scale. If we encode the value in each cell as the size of a circle rather that with a color map, we could use it as an overlay over a color heatmap for example to highlight a selected subset of the time series. As with many heatmap algorithms, bin size is a parameter to the algorithm; larger bins smooth noise and emphasize broader trends, while smaller bins help identify finer-grained phenomena.

\subsection{Implementing DenseLines on the GPU}

\begin{figure}[htb]
  \centering
  \includegraphics[width=\columnwidth]{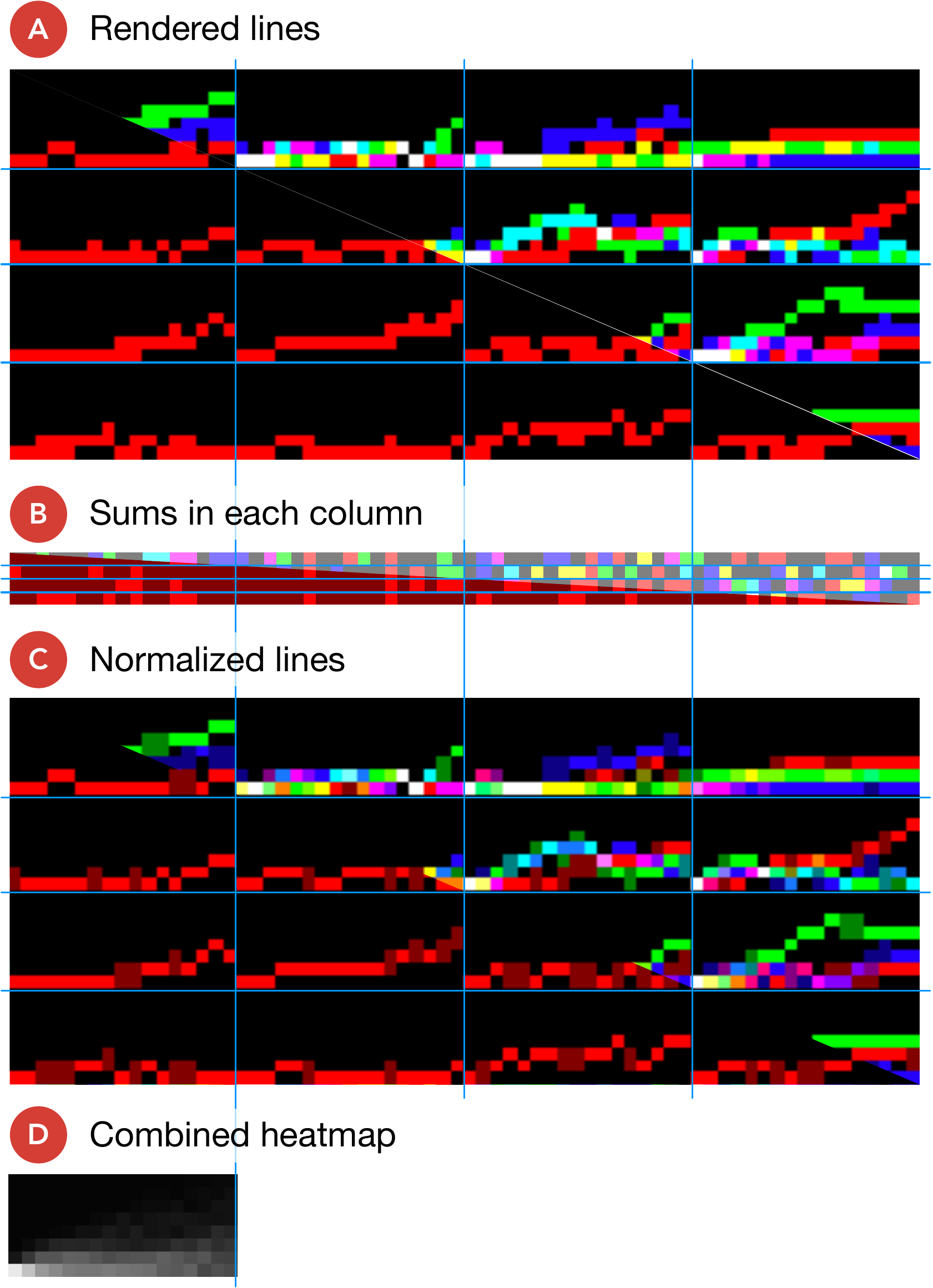}
  \vspace{-10px}
  \caption{Simplified overview of the four textures corresponding to the four steps of implementing DenseLines on a GPU. The first three images show the textures we use to exchange data between the different compute steps with red, green and blue in the upper right and---for illustration---only the red channel in the lower left. The blue grid lines show which pixels in the different textures belong to the same lines.}~\label{fig:gpu}
  \vspace{-10px}
\end{figure}

To efficiently use the GPU in our prototype JavaScript implementation, we implemented the rendering and normalization steps in WebGL shaders. \autoref{fig:gpu} gives an overview of our implementation. First, we render a batch of lines into a texture (A) of maximum size necessary and allowed by the GPU. We use the available color channels (red, green, blue and alpha) to render four lines in the same part of the texture. Lines have to be kept separate because each line needs to be normalized independently. In the second step, we compute the count of pixels in each row for each line. The result is a buffer (B) that has the same width as the texture for the lines but is only as high as there are rows of time series. In the third step, we normalize the lines in the values in the first texture (A) by the counts in the second texture (B) into a new texture (C). Lastly, we collect the normalized time series that are spread across the texture and in different color channels (C) into a single output (D). We repeat these steps until all batches of time series have been processed. You can try the demo at \url{https://domoritz.github.io/line-density}. The page has a link to the source code and the shaders.

\subsection{Limitations and Opportunities}

With large-scale data, no single technique can handle all tasks. The DenseLines technique is designed for a specific set of tasks. It is useful when there are many time series sharing the same domain and assessment of aggregate trends and outliers are more important than distinguishing the behavior of individual series. 
The technique does have some limitations. DenseLines makes it difficult, for example, to recognize information about slopes in particular areas. It is not possible to tell whether the same line is the extremum at different points in time. Some of these specific questions could be addressed with cross-highlighting, or by superimposing highlights and selections. In an interactive system, a line density visualization could be useful as part of the overview stage of information seeking~\cite{theEyesHaveIt}. In DenseLine charts the display space is binned and not continuous as in line charts. Thus, the resulting matrices can be subtracted to compute and visualize the difference between two large sets of time series.

\section{Demonstration on Public Data}

\begin{figure}[htb]
  \centering
  \includegraphics[width=.9\columnwidth]{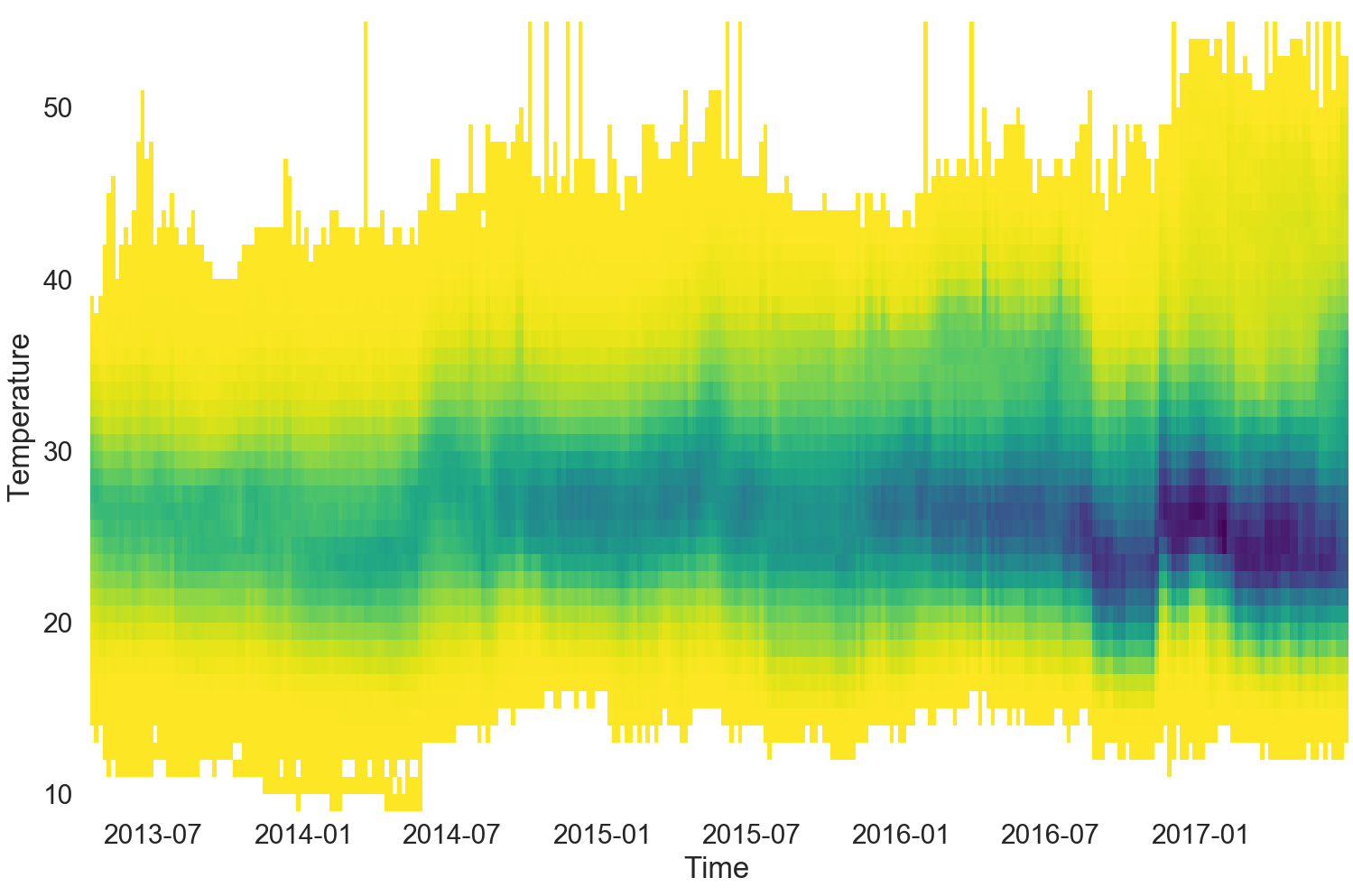}
  \vspace{-10px}
  \caption{Temperatures of 108,000 hard drives over four years.}~\label{fig:drives}
  \vspace{-10px}
\end{figure}

We first demonstrate DenseLines on a stock market dataset of $3,500$ historical New York Stock Exchange closing prices in \autoref{fig:teaser} (c). Dense clusters of lines are easy to spot in blue, while bright yellow shows areas with few stock price lines. The drop that came with the financial crisis in 2008 is clearly visible. Similarly, we can see two dense bands of stock values around $\$15$ and $\$25$, showing that companies (or customers) tend toward round stock prices.

We also examine a dataset of over $100\,\text{K}$ time series. Backblaze---a cloud storage provider with $250$\,PB of hard drive storage (Fall 2017)---publishes daily hard drive statistics from the drives in their data centers~\cite{backblaze}. \autoref{fig:drives} shows the time series of the hard drive temperature (SMART 194) for over $108,000$ hard drives. This visualization effectively displays an aggregation of $72\,\text{M}$ individual records. We can see that no hard drive goes above $55^\circ$C with most of them staying between $20^\circ$C and $30^\circ$C.

\section{Case Study: Analyzing Server Use}

\begin{figure}[htb]
  \centering
  \includegraphics[width=\columnwidth]{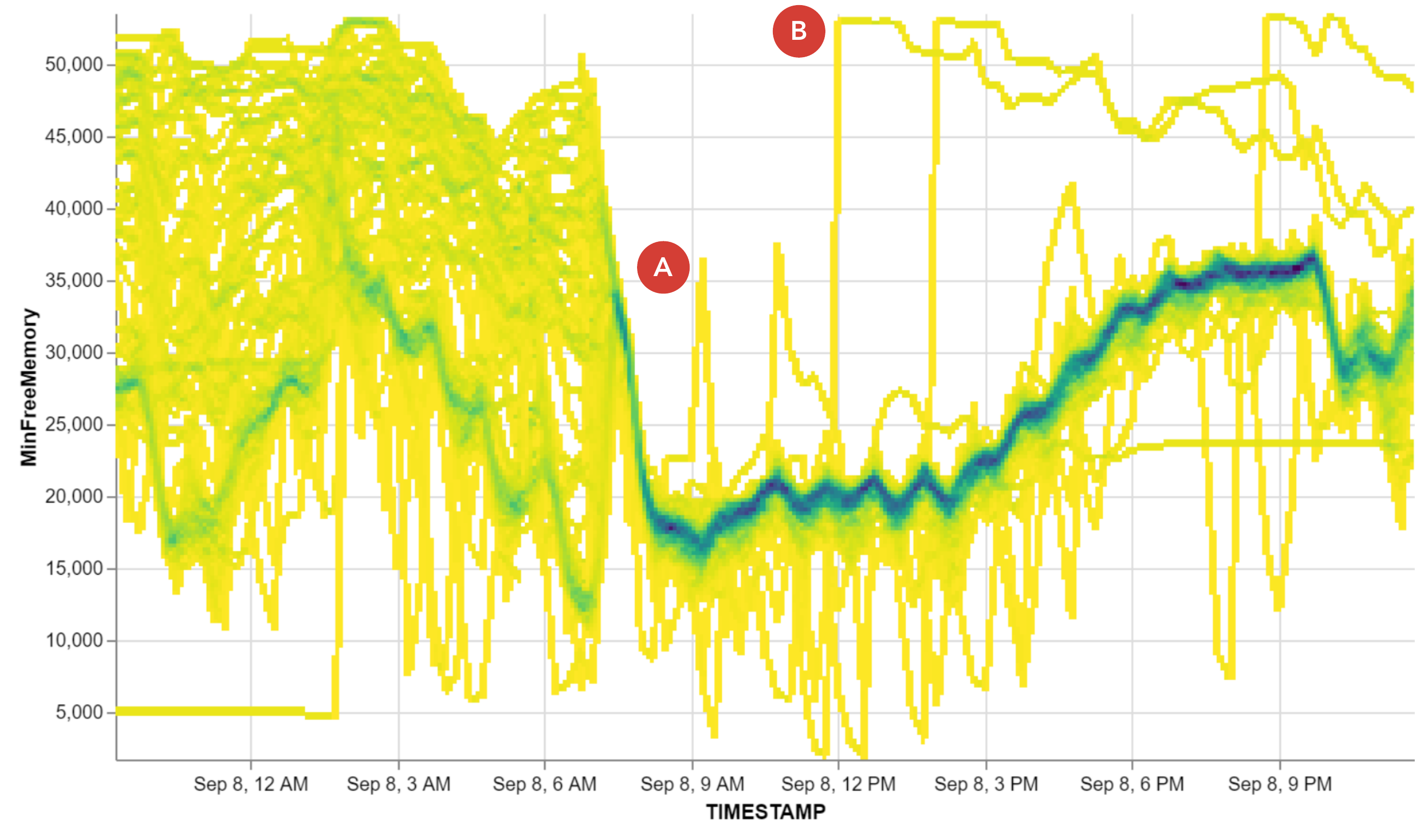}
  \vspace{-10px}
  \caption{Free memory on 140 servers over three days as a DenseLines chart. (A) On September 8, a new version was deployed and usage becomes more consistent; (B) a single server crashes .}~\label{fig:servers_heat}
  \vspace{-10px}
\end{figure}

\begin{figure}[b]
  \centering
  \includegraphics[width=\columnwidth]{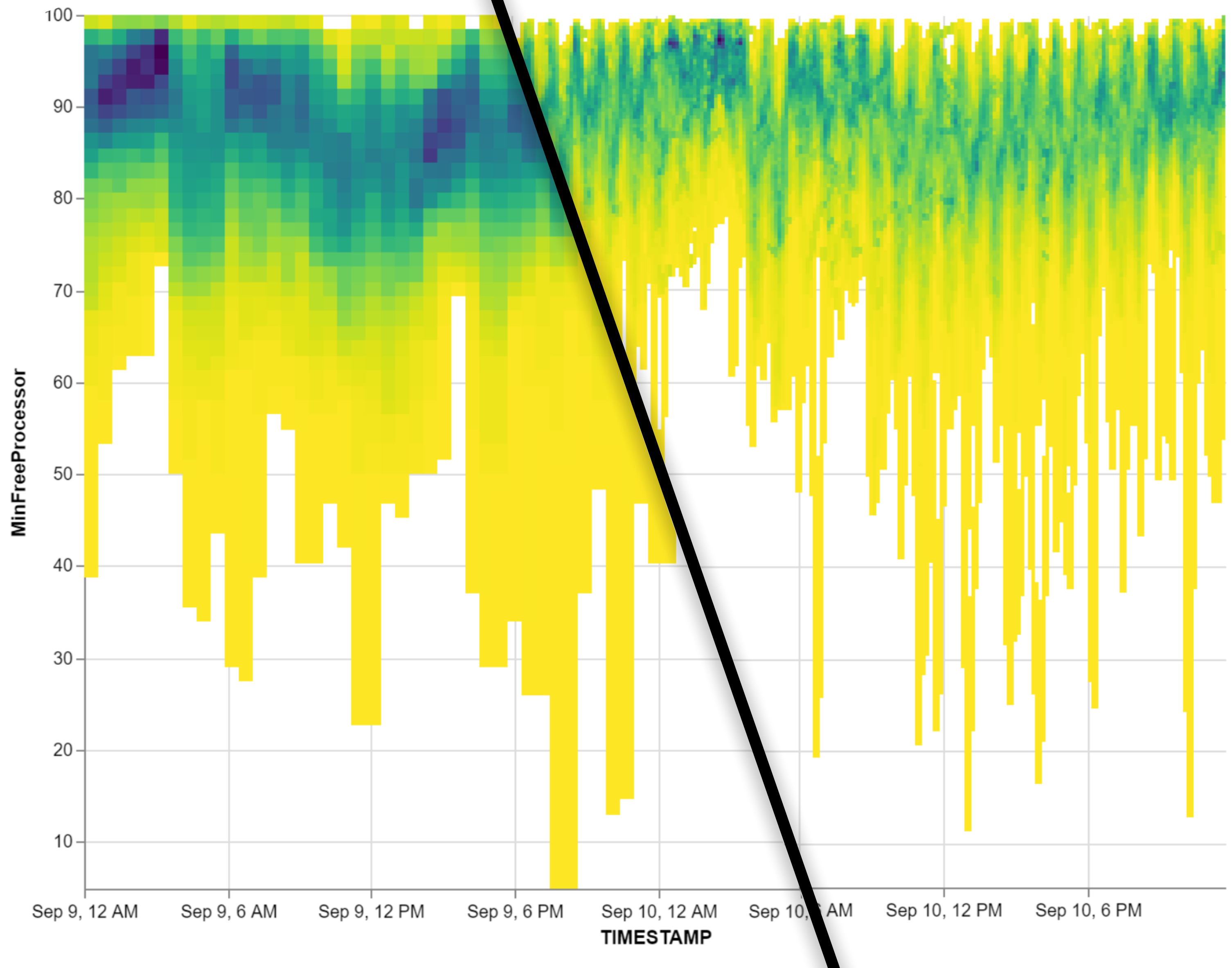}
  \vspace{-10px}
  \caption{CPU usage over time for 55 servers, using large (left) and small (right) bins. The larger three-hour bins capture the density of the space and the daily rhythm; the smaller 15-minute bins capture hourly variations.}~\label{fig:res}
  \vspace{-10px}
\end{figure}

Our case study concerns a real-life deployment of DenseLines. Brad runs operations for software-as-a-service hosted at a large cloud services organization. Among his other work, Brad is responsible for ensuring that the servers remain well-balanced. Brad was analyzing a particular cluster of 140 machines that runs a critical process. From time to time, a server would overload and crash---when it did so, it would have a great deal of free memory. The load balancer would detect this crash, restart the process, and reallocate jobs to other servers. Brad wanted to know how crashes relate to each other, and to better understand the nature of his cluster. A standard line chart of Brad's data suffers from tremendous clutter. Brad adopted a few chart variants: a chart that shows the inner percentiles of his data and another which limits the display to samples of ten lines. In both cases neither outliers nor trends were visible.

We built a version of DenseLines that works inside Brad's data analytics tool (\autoref{fig:servers_heat}). He instantly recognized the overall rhythm of the data. He pointed to the thickness of the blue line and noted that ``it shows [...] how tightly grouped things are.'' On the left side, the machines are poorly balanced; after point (A), the blue area gets dark and thin showing that the machines are well-balanced. Brad said that his analysis ``is all about outliers and deviations from the norm,'' and pointed to point (B). He recognized the distinctive pattern---a single vertical line---of a single server crashing. Brad found it helpful to see the extremes: ``The yellow shading [...] ensures that I'm seeing effectively all of the areas that have been touched by one line of some kind.'' Conversely, he found it useful to see when no machines were crashing: ``The light color is comprehensive. If it is white, there was no line that hit that.'' 

Brad showed us a different cluster of 55 servers running background tasks (\autoref{fig:res}). The servers have a slow daily cycle, running from daytime peak hours to night-time quiescence. However, they also run compute jobs every hour on the hour. Using a wide bin size, about two hours, brings out the daily cycle; a smaller bin size of fifteen minutes emphasizes the hourly spikes.

Brad has begun to incorporate DenseLines into his group's regular reviews and into his understanding of how his servers work; he presents DenseLines as part of his monitoring process.




\section{Conclusion}

DenseLines is a discrete version of Curve Density Estimates~\cite{Lampe2011} that scales well to large time series datasets.
The technique reveals the density of time series data by computing locations where multiple series share the same value. The visualization supports many typical line chart tasks, at the cost of some fidelity to individual lines. DenseLines shows places where at least one series has an outlier, and so can help locate them; it identifies dense regions and conveys the distribution of lines within these regions.
As we continue to place sensors, gather more data, and broaden analysis systems, the ability to overview and interactively explore multiple time series on similar axes will become increasingly important. We look forward to other techniques that continue to explore the space of large temporal datasets.

\begin{acks}

We thank Kim Manis, Brandon Unger, Steven Drucker, Alper Sarikaya, Ove Daae Lampe, Helwig Hauser, Carlos Scheidegger, Jeffrey Heer, Michael Correll, Matthew Conlen, and the anonymous reviewers for their comments and feedback.


\end{acks}

\bibliographystyle{ACM-Reference-Format}
\bibliography{paper}


\begin{thebibliography}{00}


\ifx \showCODEN    \undefined \def \showCODEN     #1{\unskip}     \fi
\ifx \showDOI      \undefined \def \showDOI       #1{#1}\fi
\ifx \showISBNx    \undefined \def \showISBNx     #1{\unskip}     \fi
\ifx \showISBNxiii \undefined \def \showISBNxiii  #1{\unskip}     \fi
\ifx \showISSN     \undefined \def \showISSN      #1{\unskip}     \fi
\ifx \showLCCN     \undefined \def \showLCCN      #1{\unskip}     \fi
\ifx \shownote     \undefined \def \shownote      #1{#1}          \fi
\ifx \showarticletitle \undefined \def \showarticletitle #1{#1}   \fi
\ifx \showURL      \undefined \def \showURL       {\relax}        \fi
\providecommand\bibfield[2]{#2}
\providecommand\bibinfo[2]{#2}
\providecommand\natexlab[1]{#1}
\providecommand\showeprint[2][]{arXiv:#2}

\bibitem[\protect\citeauthoryear{??}{bac}{2013}]%
        {backblaze}
 \bibinfo{year}{2013}\natexlab{}.
\newblock \bibinfo{title}{Hard Drive Data and Stats}.
\newblock   (\bibinfo{year}{2013}).
\newblock
\showURL{%
\url{https://www.backblaze.com/b2/hard-drive-test-data.html}}


\bibitem[\protect\citeauthoryear{Aigner, Miksch, Schuman, and Tominski}{Aigner
  et~al\mbox{.}}{2011}]%
        {aigner:vistime:2011}
\bibfield{author}{\bibinfo{person}{Wolfgang Aigner}, \bibinfo{person}{S.
  Miksch}, \bibinfo{person}{Heidrun Schuman}, {and} \bibinfo{person}{C.
  Tominski}.} \bibinfo{year}{2011}\natexlab{}.
\newblock \bibinfo{booktitle}{{\em Visualization of Time-Oriented Data\/}
  (\bibinfo{edition}{1st} ed.)}.
\newblock \bibinfo{publisher}{Springer Verlag}. 286 pages.
\newblock
\showISBNx{978-0-85729-078-6}
\showDOI{%
\url{https://doi.org/10.1007/978-0-85729-079-3}}


\bibitem[\protect\citeauthoryear{Artero, de~Oliveira, and Levkowitz}{Artero
  et~al\mbox{.}}{2004}]%
        {PCDensity}
\bibfield{author}{\bibinfo{person}{Almir~Olivette Artero},
  \bibinfo{person}{Maria Cristina~Ferreira de Oliveira}, {and}
  \bibinfo{person}{Haim Levkowitz}.} \bibinfo{year}{2004}\natexlab{}.
\newblock \showarticletitle{Uncovering Clusters in Crowded Parallel Coordinates
  Visualizations}. In \bibinfo{booktitle}{{\em Proceedings of the IEEE
  Symposium on Information Visualization}} {\em (\bibinfo{series}{INFOVIS
  '04})}. \bibinfo{publisher}{IEEE Computer Society},
  \bibinfo{address}{Washington, DC, USA}, \bibinfo{pages}{81--88}.
\newblock
\showISBNx{0-7803-8779-3}
\showDOI{%
\url{https://doi.org/10.1109/INFOVIS.2004.68}}


\bibitem[\protect\citeauthoryear{Bresenham}{Bresenham}{1977}]%
        {Bresenham1977}
\bibfield{author}{\bibinfo{person}{Jack Bresenham}.}
  \bibinfo{year}{1977}\natexlab{}.
\newblock \showarticletitle{{Graphics and A Linear Algorithm for Incremental
  Digital Display of Circular Arcs}}.
\newblock \bibinfo{journal}{{\em IBM System Communications Division\/}}
  \bibinfo{volume}{20(2)} (\bibinfo{year}{1977}), \bibinfo{pages}{100--106}.
\newblock
\showISSN{00010782}
\showDOI{%
\url{https://doi.org/10.1145/359423.359432}}


\bibitem[\protect\citeauthoryear{Burch, Vehlow, Beck, Diehl, and
  Weiskopf}{Burch et~al\mbox{.}}{2011}]%
        {Burch:2011:PES:2068462.2068635}
\bibfield{author}{\bibinfo{person}{Michael Burch}, \bibinfo{person}{Corinna
  Vehlow}, \bibinfo{person}{Fabian Beck}, \bibinfo{person}{Stephan Diehl},
  {and} \bibinfo{person}{Daniel Weiskopf}.} \bibinfo{year}{2011}\natexlab{}.
\newblock \showarticletitle{Parallel Edge Splatting for Scalable Dynamic Graph
  Visualization}.
\newblock \bibinfo{journal}{{\em IEEE Transactions on Visualization and
  Computer Graphics\/}} \bibinfo{volume}{17}, \bibinfo{number}{12}
  (\bibinfo{date}{Dec.} \bibinfo{year}{2011}), \bibinfo{pages}{2344--2353}.
\newblock
\showISSN{1077-2626}
\showDOI{%
\url{https://doi.org/10.1109/TVCG.2011.226}}


\bibitem[\protect\citeauthoryear{Carr, Littlefield, Nicholson, and
  Littlefield}{Carr et~al\mbox{.}}{1987}]%
        {Carr1987}
\bibfield{author}{\bibinfo{person}{D.~B. Carr}, \bibinfo{person}{R.~J.
  Littlefield}, \bibinfo{person}{W.~L. Nicholson}, {and} \bibinfo{person}{J.~S.
  Littlefield}.} \bibinfo{year}{1987}\natexlab{}.
\newblock \showarticletitle{Scatterplot Matrix Techniques for Large {N}}.
\newblock \bibinfo{journal}{{\it J. Amer. Statist. Assoc.}}
  \bibinfo{volume}{82}, \bibinfo{number}{398} (\bibinfo{year}{1987}),
  \bibinfo{pages}{424--436}.
\newblock
\showISSN{01621459}
\showDOI{%
\url{https://doi.org/10.2307/2289444}}


\bibitem[\protect\citeauthoryear{Correll, Albers, Franconeri, and
  Gleicher}{Correll et~al\mbox{.}}{2012}]%
        {comparingaverages}
\bibfield{author}{\bibinfo{person}{Michael Correll}, \bibinfo{person}{Danielle
  Albers}, \bibinfo{person}{Steven Franconeri}, {and} \bibinfo{person}{Michael
  Gleicher}.} \bibinfo{year}{2012}\natexlab{}.
\newblock \showarticletitle{Comparing Averages in Time Series Data}. In
  \bibinfo{booktitle}{{\em Proceedings of the SIGCHI Conference on Human
  Factors in Computing Systems}} {\em (\bibinfo{series}{CHI '12})}.
  \bibinfo{publisher}{ACM}, \bibinfo{address}{New York, NY, USA},
  \bibinfo{pages}{1095--1104}.
\newblock
\showISBNx{978-1-4503-1015-4}
\showDOI{%
\url{https://doi.org/10.1145/2207676.2208556}}


\bibitem[\protect\citeauthoryear{Craft~Jr}{Craft~Jr}{1970}]%
        {craft1970radio}
\bibfield{author}{\bibinfo{person}{Harold~Dumont Craft~Jr}.}
  \bibinfo{year}{1970}\natexlab{}.
\newblock \showarticletitle{Radio Observations of the Pulse Profiles and
  Dispersion Measures of Twelve Pulsars.}
\newblock  (\bibinfo{year}{1970}).
\newblock


\bibitem[\protect\citeauthoryear{Crow}{Crow}{1977}]%
        {antialias}
\bibfield{author}{\bibinfo{person}{Franklin~C. Crow}.}
  \bibinfo{year}{1977}\natexlab{}.
\newblock \showarticletitle{The Aliasing Problem in Computer-generated Shaded
  Images}.
\newblock \bibinfo{journal}{{\em Commun. ACM\/}} \bibinfo{volume}{20},
  \bibinfo{number}{11} (\bibinfo{date}{Nov.} \bibinfo{year}{1977}),
  \bibinfo{pages}{799--805}.
\newblock
\showISSN{0001-0782}
\showDOI{%
\url{https://doi.org/10.1145/359863.359869}}


\bibitem[\protect\citeauthoryear{Dean and Ghemawat}{Dean and Ghemawat}{2004}]%
        {Dean2004MapReduceSD}
\bibfield{author}{\bibinfo{person}{Jeffrey Dean} {and} \bibinfo{person}{Sanjay
  Ghemawat}.} \bibinfo{year}{2004}\natexlab{}.
\newblock \showarticletitle{MapReduce: Simplified Data Processing on Large
  Clusters}.
\newblock \bibinfo{journal}{{\em Commun. ACM\/}}  \bibinfo{volume}{51}
  (\bibinfo{year}{2004}), \bibinfo{pages}{107--113}.
\newblock
\showDOI{%
\url{https://doi.org/10.1145/1327452.1327492}}


\bibitem[\protect\citeauthoryear{Ellis and Dix}{Ellis and Dix}{2007}]%
        {clutter}
\bibfield{author}{\bibinfo{person}{G. Ellis} {and} \bibinfo{person}{A. Dix}.}
  \bibinfo{year}{2007}\natexlab{}.
\newblock \showarticletitle{A Taxonomy of Clutter Reduction for Information
  Visualisation}.
\newblock \bibinfo{journal}{{\em IEEE Transactions on Visualization and
  Computer Graphics\/}} \bibinfo{volume}{13}, \bibinfo{number}{6}
  (\bibinfo{date}{Nov} \bibinfo{year}{2007}), \bibinfo{pages}{1216--1223}.
\newblock
\showISSN{1077-2626}
\showDOI{%
\url{https://doi.org/10.1109/TVCG.2007.70535}}


\bibitem[\protect\citeauthoryear{Fulcher, Little, and Jones}{Fulcher
  et~al\mbox{.}}{2013}]%
        {Fulcher2013}
\bibfield{author}{\bibinfo{person}{Ben~D Fulcher}, \bibinfo{person}{Max~A
  Little}, {and} \bibinfo{person}{Nick~S Jones}.}
  \bibinfo{year}{2013}\natexlab{}.
\newblock \showarticletitle{{Highly comparative time-series analysis: the
  empirical structure of time series and their methods.}}
\newblock \bibinfo{journal}{{\em Journal of the Royal Society, Interface / the
  Royal Society\/}} \bibinfo{volume}{10}, \bibinfo{number}{83}
  (\bibinfo{year}{2013}).
\newblock
\showDOI{%
\url{https://doi.org/10.1098/rsif.2013.0048}}


\bibitem[\protect\citeauthoryear{Heer, Kong, and Agrawala}{Heer
  et~al\mbox{.}}{2009}]%
        {Heer2009}
\bibfield{author}{\bibinfo{person}{Jeffrey Heer}, \bibinfo{person}{Nicholas
  Kong}, {and} \bibinfo{person}{Maneesh Agrawala}.}
  \bibinfo{year}{2009}\natexlab{}.
\newblock \showarticletitle{{Sizing the horizon: the effects of chart size and
  layering on the graphical perception of time series visualizations}}.
\newblock \bibinfo{journal}{{\em CHI '09\/}} (\bibinfo{year}{2009}),
  \bibinfo{pages}{1303--1312}.
\newblock
\showISBNx{9781605582467}
\showISSN{1605582468}
\showDOI{%
\url{https://doi.org/10.1145/1518701.1518897}}


\bibitem[\protect\citeauthoryear{Heinrich and Weiskopf}{Heinrich and
  Weiskopf}{2009}]%
        {Heinrich:2009:CPC:1638611.1639151}
\bibfield{author}{\bibinfo{person}{Julian Heinrich} {and}
  \bibinfo{person}{Daniel Weiskopf}.} \bibinfo{year}{2009}\natexlab{}.
\newblock \showarticletitle{Continuous Parallel Coordinates}.
\newblock \bibinfo{journal}{{\em IEEE Transactions on Visualization and
  Computer Graphics\/}} \bibinfo{volume}{15}, \bibinfo{number}{6}
  (\bibinfo{date}{Nov.} \bibinfo{year}{2009}), \bibinfo{pages}{1531--1538}.
\newblock
\showISSN{1077-2626}
\showDOI{%
\url{https://doi.org/10.1109/TVCG.2009.131}}


\bibitem[\protect\citeauthoryear{Hinrichs, Forlini, Moynihan, Matejka,
  Anderson, and Fitzmaurice}{Hinrichs et~al\mbox{.}}{2015}]%
        {Hinrichs2015}
\bibfield{author}{\bibinfo{person}{Uta Hinrichs}, \bibinfo{person}{Stefania
  Forlini}, \bibinfo{person}{Bridget Moynihan}, \bibinfo{person}{Justin
  Matejka}, \bibinfo{person}{Fraser Anderson}, {and} \bibinfo{person}{George
  Fitzmaurice}.} \bibinfo{year}{2015}\natexlab{}.
\newblock \showarticletitle{{Dynamic Opacity Optimization for Scatter Plots}}.
\newblock \bibinfo{journal}{{\em CHI 2015\/}} (\bibinfo{year}{2015}),
  \bibinfo{pages}{2--5}.
\newblock
\showISBNx{9781450331456}
\showDOI{%
\url{https://doi.org/10.1145/2702123.2702585}}


\bibitem[\protect\citeauthoryear{Hochheiser and Shneiderman}{Hochheiser and
  Shneiderman}{2004}]%
        {Hochheiser2004}
\bibfield{author}{\bibinfo{person}{Harry Hochheiser} {and} \bibinfo{person}{Ben
  Shneiderman}.} \bibinfo{year}{2004}\natexlab{}.
\newblock \showarticletitle{{Dynamic query tools for time series data sets:
  Timebox widgets for interactive exploration}}.
\newblock \bibinfo{journal}{{\em Information Visualization\/}}
  \bibinfo{volume}{3}, \bibinfo{number}{1} (\bibinfo{year}{2004}),
  \bibinfo{pages}{1--18}.
\newblock
\showISBNx{1473-8716}
\showISSN{1473-8716}
\showDOI{%
\url{https://doi.org/10.1057/palgrave.ivs.9500061}}


\bibitem[\protect\citeauthoryear{Hyndman and Shang}{Hyndman and Shang}{2010}]%
        {functionalbagplots}
\bibfield{author}{\bibinfo{person}{Rob~J. Hyndman} {and}
  \bibinfo{person}{Han~Lin Shang}.} \bibinfo{year}{2010}\natexlab{}.
\newblock \showarticletitle{Rainbow Plots, Bagplots, and Boxplots for
  Functional Data}.
\newblock \bibinfo{journal}{{\em Journal of Computational and Graphical
  Statistics\/}} \bibinfo{volume}{19}, \bibinfo{number}{1}
  (\bibinfo{year}{2010}), \bibinfo{pages}{29--45}.
\newblock
\showDOI{%
\url{https://doi.org/10.1198/jcgs.2009.08158}}
\showeprint{https://doi.org/10.1198/jcgs.2009.08158}


\bibitem[\protect\citeauthoryear{Javed and Elmqvist}{Javed and
  Elmqvist}{2012}]%
        {6183556}
\bibfield{author}{\bibinfo{person}{W. Javed} {and} \bibinfo{person}{N.
  Elmqvist}.} \bibinfo{year}{2012}\natexlab{}.
\newblock \showarticletitle{Exploring the design space of composite
  visualization}. In \bibinfo{booktitle}{{\em 2012 IEEE Pacific Visualization
  Symposium}}. \bibinfo{pages}{1--8}.
\newblock
\showISSN{2165-8765}
\showDOI{%
\url{https://doi.org/10.1109/PacificVis.2012.6183556}}


\bibitem[\protect\citeauthoryear{Javed, McDonnel, and Elmqvist}{Javed
  et~al\mbox{.}}{2010}]%
        {Javed2010}
\bibfield{author}{\bibinfo{person}{Waqas Javed}, \bibinfo{person}{Bryan
  McDonnel}, {and} \bibinfo{person}{Niklas Elmqvist}.}
  \bibinfo{year}{2010}\natexlab{}.
\newblock \showarticletitle{{Graphical perception of multiple time series}}.
\newblock \bibinfo{journal}{{\em IEEE Transactions on Visualization and
  Computer Graphics\/}} \bibinfo{volume}{16}, \bibinfo{number}{6}
  (\bibinfo{year}{2010}), \bibinfo{pages}{927--934}.
\newblock
\showISBNx{1077-2626}
\showISSN{10772626}
\showDOI{%
\url{https://doi.org/10.1109/TVCG.2010.162}}


\bibitem[\protect\citeauthoryear{Jugel, Jerzak, Hackenbroich, and Markl}{Jugel
  et~al\mbox{.}}{2014}]%
        {Jugel2014M4AV}
\bibfield{author}{\bibinfo{person}{Uwe Jugel}, \bibinfo{person}{Zbigniew
  Jerzak}, \bibinfo{person}{Gregor Hackenbroich}, {and} \bibinfo{person}{Volker
  Markl}.} \bibinfo{year}{2014}\natexlab{}.
\newblock \showarticletitle{M4: A Visualization-Oriented Time Series Data
  Aggregation}.
\newblock \bibinfo{journal}{{\em PVLDB\/}}  \bibinfo{volume}{7}
  (\bibinfo{year}{2014}), \bibinfo{pages}{797--808}.
\newblock
\showDOI{%
\url{https://doi.org/10.14778/2732951.2732953}}


\bibitem[\protect\citeauthoryear{Kandel, Parikh, Paepcke, Hellerstein, and
  Heer}{Kandel et~al\mbox{.}}{2012}]%
        {Kandel2012}
\bibfield{author}{\bibinfo{person}{Sean Kandel}, \bibinfo{person}{Ravi Parikh},
  \bibinfo{person}{Andreas Paepcke}, \bibinfo{person}{Joseph~M Hellerstein},
  {and} \bibinfo{person}{Jeffrey Heer}.} \bibinfo{year}{2012}\natexlab{}.
\newblock \showarticletitle{{Profiler : Integrated Statistical Analysis and
  Visualization for Data Quality Assessment}}.
\newblock \bibinfo{journal}{{\em Proceedings of Advanced Visual Interfaces,
  AVI\/}} (\bibinfo{year}{2012}), \bibinfo{pages}{547--554}.
\newblock
\showISBNx{9781450312875}
\showISSN{{\textless}null{\textgreater}}
\showDOI{%
\url{https://doi.org/10.1145/2254556.2254659}}
\showeprint{10.1145/2254556.2254659}


\bibitem[\protect\citeauthoryear{Kincaid and Lam}{Kincaid and Lam}{2006}]%
        {Kincaid2006}
\bibfield{author}{\bibinfo{person}{Robert Kincaid} {and} \bibinfo{person}{Heidi
  Lam}.} \bibinfo{year}{2006}\natexlab{}.
\newblock \showarticletitle{{Line Graph Explorer: scalable display of line
  graphs using Focus+Context}}.
\newblock \bibinfo{journal}{{\em AVI '06: Proceedings of the Working Conference
  on Advanced Visual Interfaces\/}} (\bibinfo{year}{2006}),
  \bibinfo{pages}{404--411}.
\newblock
\showISBNx{1595933530}
\showDOI{%
\url{https://doi.org/10.1145/1133265.1133348}}


\bibitem[\protect\citeauthoryear{Konyha, Le\v{z}, Matkovi\'{c}, Jelovi\'{c},
  and Hauser}{Konyha et~al\mbox{.}}{2012}]%
        {Konyha2012}
\bibfield{author}{\bibinfo{person}{Zolt\'{a}n Konyha}, \bibinfo{person}{Alan
  Le\v{z}}, \bibinfo{person}{Kre\v{s}imir Matkovi\'{c}}, \bibinfo{person}{Mario
  Jelovi\'{c}}, {and} \bibinfo{person}{Helwig Hauser}.}
  \bibinfo{year}{2012}\natexlab{}.
\newblock \showarticletitle{Interactive Visual Analysis of Families of Curves
  Using Data Aggregation and Derivation}. In \bibinfo{booktitle}{{\em
  Proceedings of the 12th International Conference on Knowledge Management and
  Knowledge Technologies}} {\em (\bibinfo{series}{i-KNOW '12})}.
  \bibinfo{publisher}{ACM}, \bibinfo{address}{New York, NY, USA}, Article
  \bibinfo{articleno}{24}, \bibinfo{numpages}{8}~pages.
\newblock
\showISBNx{978-1-4503-1242-4}
\showDOI{%
\url{https://doi.org/10.1145/2362456.2362487}}


\bibitem[\protect\citeauthoryear{Lampe and Hauser}{Lampe and Hauser}{2011}]%
        {Lampe2011}
\bibfield{author}{\bibinfo{person}{O.~Daae Lampe} {and} \bibinfo{person}{H.
  Hauser}.} \bibinfo{year}{2011}\natexlab{}.
\newblock \showarticletitle{Curve Density Estimates}.
\newblock \bibinfo{journal}{{\em Computer Graphics Forum\/}}
  \bibinfo{volume}{30}, \bibinfo{number}{3} (\bibinfo{date}{6}
  \bibinfo{year}{2011}), \bibinfo{pages}{633--642}.
\newblock
\showDOI{%
\url{https://doi.org/10.1111/j.1467-8659.2011.01912.x}}


\bibitem[\protect\citeauthoryear{McLachlan, Munzner, Koutsofios, and
  North}{McLachlan et~al\mbox{.}}{2008}]%
        {McLachlan2008}
\bibfield{author}{\bibinfo{person}{Peter McLachlan}, \bibinfo{person}{Tamara
  Munzner}, \bibinfo{person}{Eleftherios Koutsofios}, {and}
  \bibinfo{person}{Stephen North}.} \bibinfo{year}{2008}\natexlab{}.
\newblock \showarticletitle{{LiveRAC: Interactive Visual Exploration of System
  Management Time-Series Data}}.
\newblock \bibinfo{journal}{{\em Human Factors\/}} (\bibinfo{year}{2008}),
  \bibinfo{pages}{1483--1492}.
\newblock
\showISBNx{9781605580111}
\showDOI{%
\url{https://doi.org/10.1145/1357054.1357286}}


\bibitem[\protect\citeauthoryear{Meyer, Scheidegger, Schreiner, Duffy, Carr,
  and Silva}{Meyer et~al\mbox{.}}{2008}]%
        {Scheideggernormalize}
\bibfield{author}{\bibinfo{person}{M. Meyer}, \bibinfo{person}{C.~E.
  Scheidegger}, \bibinfo{person}{J.~M. Schreiner}, \bibinfo{person}{B. Duffy},
  \bibinfo{person}{H. Carr}, {and} \bibinfo{person}{C.~T. Silva}.}
  \bibinfo{year}{2008}\natexlab{}.
\newblock \showarticletitle{Revisiting Histograms and Isosurface Statistics}.
\newblock \bibinfo{journal}{{\em IEEE Transactions on Visualization and
  Computer Graphics\/}} \bibinfo{volume}{14}, \bibinfo{number}{6}
  (\bibinfo{date}{Nov} \bibinfo{year}{2008}), \bibinfo{pages}{1659--1666}.
\newblock
\showISSN{1077-2626}
\showDOI{%
\url{https://doi.org/10.1109/TVCG.2008.160}}


\bibitem[\protect\citeauthoryear{Playfair}{Playfair}{1801}]%
        {playfair1801commercial}
\bibfield{author}{\bibinfo{person}{William Playfair}.}
  \bibinfo{year}{1801}\natexlab{}.
\newblock \bibinfo{booktitle}{{\em The commercial and political atlas:
  representing, by means of stained copper-plate charts, the progress of the
  commerce, revenues, expenditure and debts of england during the whole of the
  eighteenth century}}.
\newblock \bibinfo{publisher}{T. Burton}.
\newblock


\bibitem[\protect\citeauthoryear{Saito, Miyamura, Yamamoto, Saito, Hoshiya, and
  Kaseda}{Saito et~al\mbox{.}}{2005}]%
        {Saito2005}
\bibfield{author}{\bibinfo{person}{Takafumi Saito},
  \bibinfo{person}{Hiroko~Nakamura Miyamura}, \bibinfo{person}{Mitsuyoshi
  Yamamoto}, \bibinfo{person}{Hiroki Saito}, \bibinfo{person}{Yuka Hoshiya},
  {and} \bibinfo{person}{Takumi Kaseda}.} \bibinfo{year}{2005}\natexlab{}.
\newblock \showarticletitle{{Two-tone pseudo coloring: Compact visualization
  for one-dimensional data}}. In \bibinfo{booktitle}{{\em Proceedings - IEEE
  Symposium on Information Visualization, INFO VIS}}.
  \bibinfo{pages}{173--180}.
\newblock
\showISBNx{078039464X}
\showISSN{1522404X}
\showDOI{%
\url{https://doi.org/10.1109/INFVIS.2005.1532144}}


\bibitem[\protect\citeauthoryear{Scheepens, Willems, van~de Wetering,
  Andrienko, Andrienko, and van Wijk}{Scheepens et~al\mbox{.}}{2011}]%
        {Scheepens}
\bibfield{author}{\bibinfo{person}{Roeland Scheepens}, \bibinfo{person}{Niels
  Willems}, \bibinfo{person}{Huub van~de Wetering}, \bibinfo{person}{Gennady
  Andrienko}, \bibinfo{person}{Natalia Andrienko}, {and}
  \bibinfo{person}{Jarke~J. van Wijk}.} \bibinfo{year}{2011}\natexlab{}.
\newblock \showarticletitle{Composite Density Maps for Multivariate
  Trajectories}.
\newblock \bibinfo{journal}{{\em IEEE Transactions on Visualization and
  Computer Graphics\/}} \bibinfo{volume}{17}, \bibinfo{number}{12}
  (\bibinfo{date}{Dec.} \bibinfo{year}{2011}), \bibinfo{pages}{2518--2527}.
\newblock
\showISSN{1077-2626}
\showDOI{%
\url{https://doi.org/10.1109/TVCG.2011.181}}


\bibitem[\protect\citeauthoryear{Shneiderman}{Shneiderman}{1996}]%
        {theEyesHaveIt}
\bibfield{author}{\bibinfo{person}{B. Shneiderman}.}
  \bibinfo{year}{1996}\natexlab{}.
\newblock \showarticletitle{The eyes have it: a task by data type taxonomy for
  information visualizations}. In \bibinfo{booktitle}{{\em Proceedings 1996
  IEEE Symposium on Visual Languages}}. \bibinfo{pages}{336--343}.
\newblock
\showISSN{1049-2615}
\showDOI{%
\url{https://doi.org/10.1109/VL.1996.545307}}


\bibitem[\protect\citeauthoryear{Simkin and Hastie}{Simkin and Hastie}{1987}]%
        {Simkin1987}
\bibfield{author}{\bibinfo{person}{David Simkin} {and} \bibinfo{person}{Reid
  Hastie}.} \bibinfo{year}{1987}\natexlab{}.
\newblock \showarticletitle{{An Information-Processing Analysis of Graph
  Perception}}.
\newblock \bibinfo{journal}{{\em Source Journal of the American Statistical
  Association\/}} \bibinfo{volume}{82}, \bibinfo{number}{398}
  (\bibinfo{year}{1987}), \bibinfo{pages}{454--465}.
\newblock
\showISBNx{0471109916}
\showISSN{0162-1459}
\showDOI{%
\url{https://doi.org/10.1080/01621459.1987.10478448}}


\bibitem[\protect\citeauthoryear{Smith and van~der Walt}{Smith and van~der
  Walt}{2015}]%
        {viridis}
\bibfield{author}{\bibinfo{person}{Nathaniel Smith} {and}
  \bibinfo{person}{St\'{e}fan van~der Walt}.} \bibinfo{year}{2015}\natexlab{}.
\newblock \bibinfo{title}{A Better Default Colormap for Matplotlib}.
\newblock   (\bibinfo{year}{2015}).
\newblock
\showURL{%
\url{https://www.youtube.com/watch?v=xAoljeRJ3lU}}


\bibitem[\protect\citeauthoryear{Swihart, Caffo, James, Strand, Schwartz, and
  Punjabi}{Swihart et~al\mbox{.}}{2010}]%
        {Swihart2010}
\bibfield{author}{\bibinfo{person}{Bruce~J Swihart}, \bibinfo{person}{Brian
  Caffo}, \bibinfo{person}{Bryan~D James}, \bibinfo{person}{Matthew Strand},
  \bibinfo{person}{Brian~S Schwartz}, {and} \bibinfo{person}{Naresh~M
  Punjabi}.} \bibinfo{year}{2010}\natexlab{}.
\newblock \showarticletitle{{Lasagna plots: a saucy alternative to spaghetti
  plots.}}
\newblock \bibinfo{journal}{{\em Epidemiology (Cambridge, Mass.)\/}}
  \bibinfo{volume}{21}, \bibinfo{number}{5} (\bibinfo{year}{2010}),
  \bibinfo{pages}{621--5}.
\newblock
\showISBNx{1531-5487; 1044-3983}
\showISSN{1531-5487}
\showDOI{%
\url{https://doi.org/10.1097/EDE.0b013e3181e5b06a}}


\bibitem[\protect\citeauthoryear{Talbot, Gerth, and Hanrahan}{Talbot
  et~al\mbox{.}}{2011}]%
        {Talbot2011ArcLA}
\bibfield{author}{\bibinfo{person}{Justin Talbot}, \bibinfo{person}{John
  Gerth}, {and} \bibinfo{person}{Pat Hanrahan}.}
  \bibinfo{year}{2011}\natexlab{}.
\newblock \showarticletitle{Arc Length-Based Aspect Ratio Selection}.
\newblock \bibinfo{journal}{{\em IEEE Transactions on Visualization and
  Computer Graphics\/}}  \bibinfo{volume}{17} (\bibinfo{year}{2011}),
  \bibinfo{pages}{2276--2282}.
\newblock
\showDOI{%
\url{https://doi.org/10.1109/TVCG.2011.167}}


\bibitem[\protect\citeauthoryear{Tufte and Schmieg}{Tufte and Schmieg}{1985}]%
        {tufte1985visual}
\bibfield{author}{\bibinfo{person}{Edward~R Tufte} {and}
  \bibinfo{person}{Glenn~M Schmieg}.} \bibinfo{year}{1985}\natexlab{}.
\newblock \showarticletitle{The visual display of quantitative information}.
\newblock \bibinfo{journal}{{\em American Journal of Physics\/}}
  \bibinfo{volume}{53}, \bibinfo{number}{11} (\bibinfo{year}{1985}),
  \bibinfo{pages}{1117--1118}.
\newblock


\bibitem[\protect\citeauthoryear{Wattenberg}{Wattenberg}{2001}]%
        {Wattenberg:2001:SGQ:634067.634292}
\bibfield{author}{\bibinfo{person}{Martin Wattenberg}.}
  \bibinfo{year}{2001}\natexlab{}.
\newblock \showarticletitle{Sketching a Graph to Query a Time-series Database}.
  In \bibinfo{booktitle}{{\em CHI '01 Extended Abstracts on Human Factors in
  Computing Systems}} {\em (\bibinfo{series}{CHI EA '01})}.
  \bibinfo{publisher}{ACM}, \bibinfo{address}{New York, NY, USA},
  \bibinfo{pages}{381--382}.
\newblock
\showISBNx{1-58113-340-5}
\showDOI{%
\url{https://doi.org/10.1145/634067.634292}}


\bibitem[\protect\citeauthoryear{Wickham}{Wickham}{2013}]%
        {Wickham2013}
\bibfield{author}{\bibinfo{person}{Hadley Wickham}.}
  \bibinfo{year}{2013}\natexlab{}.
\newblock \showarticletitle{{Bin-summarise-smooth : A framework for visualising
  large data}}.
\newblock \bibinfo{journal}{{\em InfoVis 2013\/}} \bibinfo{number}{August}
  (\bibinfo{year}{2013}).
\newblock
\showURL{%
\url{http://vita.had.co.nz/papers/bigvis.html}}


\bibitem[\protect\citeauthoryear{Zinsmaier, Brandes, Deussen, and
  Strobelt}{Zinsmaier et~al\mbox{.}}{2012}]%
        {zinsmaier2012interactive}
\bibfield{author}{\bibinfo{person}{Michael Zinsmaier}, \bibinfo{person}{Ulrik
  Brandes}, \bibinfo{person}{Oliver Deussen}, {and} \bibinfo{person}{Hendrik
  Strobelt}.} \bibinfo{year}{2012}\natexlab{}.
\newblock \showarticletitle{Interactive level-of-detail rendering of large
  graphs}.
\newblock \bibinfo{journal}{{\em IEEE Transactions on Visualization and
  Computer Graphics\/}} \bibinfo{volume}{18}, \bibinfo{number}{12}
  (\bibinfo{year}{2012}), \bibinfo{pages}{2486--2495}.
\newblock
\showDOI{%
\url{https://doi.org/10.1109/TVCG.2012.238}}


\end{thebibliography}

\end{document}